\def\year{2021}\relax
\documentclass[letterpaper]{article} 
\pdfoutput=1
\usepackage{aaai21}  
\usepackage{times}  
\usepackage{helvet} 
\usepackage{courier}  
\usepackage[hyphens]{url}  
\usepackage{graphicx} 
\usepackage{graphicx}  
\usepackage{natbib}  
\usepackage{balance}
\usepackage{microtype}
\usepackage{flushend}
\usepackage{booktabs} 
\usepackage{caption}
\usepackage{verbatim}
\usepackage{amsmath,amssymb,amsfonts}
\usepackage{graphicx}
\usepackage{textcomp}
\usepackage{xcolor}
\usepackage{booktabs} 
\usepackage{url}
\usepackage{threeparttable}
\usepackage{array}
\usepackage{capt-of}
\usepackage{balance,bm}
\usepackage{color, colortbl, xcolor}
\usepackage [english]{babel}
\usepackage{url}
\usepackage{graphicx}  
\usepackage{textcomp, soul}
\usepackage{tabularx}
\usepackage{longtable,booktabs}
\usepackage{wrapfig}
\usepackage{subcaption}
\usepackage{caption}
\usepackage{amsmath}
\usepackage{mathtools}
\usepackage{color,soul}
\usepackage{algorithmicx}
\usepackage[ruled]{algorithm2e}
\usepackage[T1]{fontenc}
\usepackage[utf8]{inputenc}
\usepackage{array}
\usepackage{graphicx, color, colortbl, balance, url, booktabs, soul, array, multirow, amsmath, verbatim, xspace, tabularx, wrapfig}
\mathchardef\mhyphen="2D 

\usepackage [autostyle, english = american]{csquotes}
\definecolor{linkColor}{RGB}{6,125,233}
\definecolor{green}{rgb}{0.0, 0.65, 0.31}
\definecolor{bleudefrance}{rgb}{0.19, 0.55, 0.91}
\definecolor{ceruleanblue}{rgb}{0.16, 0.32, 0.75}
\definecolor{grey}{HTML}{969696}
\definecolor{violet}{HTML}{6a51a3}
\definecolor{dgrey}{HTML}{01665e}
\definecolor{lgrey}{HTML}{5ab4ac}
\definecolor{dgreen}{HTML}{005a32}
\definecolor{purple}{HTML}{ae017e}
\definecolor{orange}{HTML}{d95f0e}

\definecolor{editCol}{rgb}{0.1, 0.1, 0.8}
\definecolor{maskCol}{HTML}{c51b7d}
\definecolor{lrColor}{HTML}{8856a7}
\definecolor{trColor}{HTML}{d01c8b}
\definecolor{ctColor}{HTML}{4dac26}
\definecolor{brickred}{HTML}{f03b20}
\definecolor{improveCol}{HTML}{253494}
\definecolor{worsenCol}{HTML}{d7191c}
\definecolor{neutralCol}{HTML}{dd1c77}
\definecolor{neutralGreen}{HTML}{31a354}
\definecolor{bleudefrance}{rgb}{0.19, 0.55, 0.91}  
\definecolor{mediumblue}{rgb}{0.0, 0.0, 0.8}
\definecolor{lgrey}{HTML}{5ab4ac}
\definecolor{dgreen}{HTML}{005a32}
\definecolor{brightgreen}{HTML}{31a354}
\definecolor{purple}{HTML}{ae017e}
\definecolor{lblue}{HTML}{decbe4}
\definecolor{deepgrey}{HTML}{525252}
\definecolor{dslate}{HTML}{2F4F4F}
\definecolor{dolive}{HTML}{556B2F}
\definecolor{teal}{HTML}{388E8E}
\definecolor{mscolor}{HTML}{4dac26}
\definecolor{nmscolor}{HTML}{d01c8b}
\definecolor{NewBlue}{HTML}{1879ba}
\definecolor{DarkBlue}{HTML}{00008B}

\usepackage{mathtools}

\usepackage{amsmath, amssymb}
\usepackage{array}
\usepackage{xcolor}
\usepackage{arydshln}
\newcommand*{\textlabel}[2]{%
  \edef\@currentlabel{#1}
  \phantomsection
  #1\label{#2}
}

\colorlet{tableheadcolor}{gray!25} 
\colorlet{tablerowcolor}{gray!15} 
\colorlet{tablerowcolor2}{gray!12} 
\colorlet{tablerowcolor3}{gray!25} 

\newcommand{\rowcollight}{\rowcolor{tablerowcolor2}} %

\colorlet{tableheadcolor}{gray!25} 
\colorlet{tablerowcolor}{gray!5} 

\definecolor{neutralCol}{HTML}{dd1c77}
\definecolor{neutralGreen}{HTML}{31a354}
\definecolor{NewBlue}{HTML}{1879ba}
\definecolor{bleudefrance}{rgb}{0.19, 0.55, 0.91}  
\definecolor{AfTrColor}{HTML}{0868ac}  
\definecolor{BfTrColor}{HTML}{a8ddb5}  

\definecolor{AfCtColor}{HTML}{b10026}  
\definecolor{BfCtColor}{HTML}{fd8d3c}

\graphicspath{ {figures/} }

 \definecolor{TrmtColor}{RGB}{47, 79, 79}
\definecolor{CtrlColor}{RGB}{162, 205, 90}
\definecolor{edit}{rgb}{0.0, 0.0, 0.0}

\usepackage{caption}
\usepackage{booktabs}
\usepackage{blindtext}

\newcommand{\Cs}{\textit{Suspended }}
\newcommand{\Ct}{\textit{Control }}
\newif{\ifhidecomments}
  \hidecommentsfalse 
\ifhidecomments
    \newcommand{\farhan}[1]{}
    \newcommand{\koustuv}[1]{}
    \newcommand{\mueen}[1]{}
\else
    \newcommand{\farhan}[1]{\textbf{\small\sffamily{\textcolor{blue}{[#1 -- Farhan]}}}}
    \newcommand{\koustuv}[1]{\textbf{\small\sffamily{\textcolor{green}{[#1 -- Koustuv]}}}}
    \newcommand{\mueen}[1]{\textbf{\small\sffamily{\textcolor{red}{[#1 -- Mueen]}}}}
\fi

\title{Examining Factors Associated with Twitter Account Suspension Following the 2020 U.S. Presidential Election}
\author{
    \fontsize{11.5}{11.5}\selectfont
    Farhan Asif Chowdhury\textsuperscript{\rm 1},
    Dheeman Saha\textsuperscript{\rm 1},
    Md Rashidul Hasan\textsuperscript{\rm 1},
    Koustuv Saha\textsuperscript{\rm 2},
    Abdullah Mueen\textsuperscript{\rm 1}
    \\
}
\affiliations{
    \textsuperscript{\rm 1}University of New Mexico,
    \textsuperscript{\rm 2}Georgia Tech
    
    \{fasifchowdhury, dsaha, mdhasan, mueen\}@unm.edu, koustuv.saha@gatech.edu
}

\begin{document}

\maketitle

\begin{abstract}
Online social media enables mass-level, transparent, and democratized discussion on numerous socio-political issues. Due to such openness, these platforms often endure manipulation and misinformation - leading to negative impacts. To prevent such harmful activities, platform moderators employ countermeasures to safeguard against actors violating their rules. However, the correlation between publicly outlined policies and employed action is less clear to general people.  

In this work, we examine violations and subsequent moderations related to the 2020 U.S. President Election discussion on Twitter. We focus on quantifying plausible reasons for the suspension, drawing on Twitter's rules and policies by identifying suspended users (\textsl{Case}) and comparing their activities and properties with (yet) non-suspended (\textsl{Control}) users. Using a dataset of 240M election-related tweets made by 21M unique users, we observe that \textsl{Suspended} users violate Twitter's rules at a higher rate (statistically significant) than \textsl{Control} users across all the considered aspects - hate speech, offensiveness, spamming, and civic integrity. Moreover, through the lens of Twitter's suspension mechanism, we qualitatively examine the targeted topics for manipulation.

\end{abstract}

\section{Introduction}
Social media platforms such as Facebook, Twitter, and Reddit have become vastly popular in public discussion related to societal, economic, and political issues~\cite{gil2012social}. However, in the recent past, these platforms were heavily targeted for manipulation and spreading misinformation relating to numerous civic issues across the world~\cite{bessi2016social}, which is often referred to as ``Computational Propaganda''~\cite{woolley2018computational}. In particular, the coordinated misinformation and influence campaigns of foreign state-sponsored actors during the 2016 U.S. presidential election were highly scrutinized - which eventually led to the U.S. Congressional hearings and investigation by the U.S. Department of Justice \cite{congress_hearing, mueller_report}.



In the aftermath of the 2016 U.S. presidential election, social platforms announced strict and improved platform moderation policy~\cite{facebook_update,twitter_update}. However, these platform's moderation and suspension policies have been largely debated and have faced severe criticism from the political leaders and supporters about bias towards their opposition~\cite{twitter_bias_1, twitter_bias_2}. Although these platforms publicly outline their moderation policy, there is no third-party monitoring of their enacted moderation. Moreover, social platforms like Twitter and Facebook employ extensive safeguard mechanisms that consider various aspects of user activities (coordinated activities, impersonation, etc.) to identify malicious behavior~\cite{twitter_safety}. Therefore, analyzing these suspended users' tweets and shared content might shed light on violators' targeted topics.


In the context of inadequate countermeasures against manipulation during the 2016 U.S. presidential election, the 2020 election was of paramount importance for platform operators to provide a safe and democratized public discussion sphere~\cite{twitter_election_policy}. The impact and importance of these safeguard measures are not confined to this particular election; instead, they bear cardinal implications for future online political discussions exceeding all geopolitical boundaries. Therefore, it is requisite to assess these platforms' moderation policy --- to investigate the correlation between their policies and actions, examine for potential political biases, and make general people aware of the targeted malice topics.

In this respect, we focus on analyzing the moderation policy of popular micro-blogging site Twitter as a case-study by asking the following research questions:
\begin{itemize}
    \item \textbf{RQ1:} What factors associate with Twitter account suspension following the 2020 U.S. President Election?
    \item \textbf{RQ2:} How do political ideologies associate with the suspended accounts?
    \item \textbf{RQ3:} What was the topic of discussion among suspended users? What type of content these users shared? 
\end{itemize}



\noindent\textbf{This work.} To answer these research questions, we collect a large-scale dataset of 240M tweets made by 21M unique users over eight weeks centering the 2020 U.S. President Election. Afterward, we identify 355K suspended users who participated in this election discussion. We draw upon Twitter's rules and policies to examine plausible suspension factors. To investigate the user activity that might lead to suspension, we adopt the \textsl{``case-control''} study design from epidemiology~\cite{schulz2002case}. We consider the suspended users as \textsl{Case} group and sample similar number of non-suspended users as \textsl{Control} group. We devise several classification techniques to quantify suspension factors among these two groups. We infer these users' political leaning by utilizing the political bias of the news media outlet they share. By employing a language differentiation technique, we contrast the conversational topics and shared content among \textsl{Case} and \textsl{Control} groups. Through the lens of Twitter's suspension policy, we passively infer the targeted topics by platform violators and identify the online content platforms utilized to sway the discussion.

\noindent\textbf{Summary findings.} We find that across all suspension factors, the \textsl{Suspended} users have higher (statistically significant) violation occurrences. Coherent with prior work, we find that \textsl{Suspended} users are short-lived, have fewer followers, and show more tweeting activity. We observe a higher presence of right-leaning users than left-leaning users among \textsl{Suspended} users. We find that \textsl{Suspended} users use more curse and derogatory words and personally-attacking and propaganda related hashtags. We also notice that these users share news content from heavily biased right-leaning news-propaganda sites. We discuss the implications and limitations of our work in the Conclusion.      





\section{Related Work}
There have been several works related to Twitter suspension, most of which focused primarily on spam-related suspensions~\cite{thomas2011suspended, amleshwaram2013cats}. More recently,~\citeauthor{le2019postmortem} studied suspended users in the context of the 2016 U.S. presidential election~\cite{le2019postmortem}, and~\citeauthor{chowdhury2020twitter} examined a large group of suspended users related to a large-scale Twitter purge in 2018~\cite{chowdhury2020twitter}. We refer readers to ~\cite{le2019postmortem,chowdhury2020twitter}  for a more comprehensive understanding of suspension and moderation on online platforms. However, none of these works quantify specific factors associated with Twitter suspension. Additionally, political discussions and related manipulation on online platforms have been thoroughly studied previously~\cite{ferrara2017disinformation, im2020still}, mostly related to the 2016 U.S.presidential election~\cite{badawy2018analyzing, zannettou2019disinformation}. These works primarily focus on characterizing malicious users and inferring their motivation and impact.~\citeauthor{im2020still, badawy2018analyzing} provide an extensive overview of this line of work~\cite{im2020still, badawy2018analyzing}. In contrast, we focus on quantifying suspension factors and examining malice topics related to the 2020 U.S. presidential election.    

\section{Data}
To collect tweets related to the 2020 U.S. President Election discussion, we deployed an uninterrupted data collection framework utilizing Twitter's streaming API to filter real-time tweets based on given keywords. Similar to prior work that collected specific theme or event-related tweets  \cite{olteanu2014crisislex}, we initialize the keyword set with manually curated election-related words and hashtags. To cover the continuously evolving election discussions and topics, we update the keyword set daily with new trending hashtags and words from previous days collection, as employed in \cite{abu2019botcamp, olteanu2015comparing}. We provide a list of all the keywords in the supporting website \cite{support_materials_website}. Our data collection time-period spans over around ten weeks, centering the election date - from ``September 28, 2020'' to ``December 04, 2020''. Approximately one month after our data collection ends, on ``January 01, 2021'' we start probing Twitter for each of the participating users from our dataset to identify the suspended users. Twitter returns the response code $63$ when requested user information for a suspended user. In this process, we identify 355,573 suspended users from roughly 21M participating users. We provide a summary descriptive statistics of our dataset in Table 1 and plot per user tweet count in Figure 1. We discuss the limitation of our data collection framework in Section 6.\looseness=-1

\begin{table}[t]
\sffamily
\footnotesize 
  \begin{minipage}[t]{0.42\columnwidth}
    \includegraphics[width=\columnwidth]{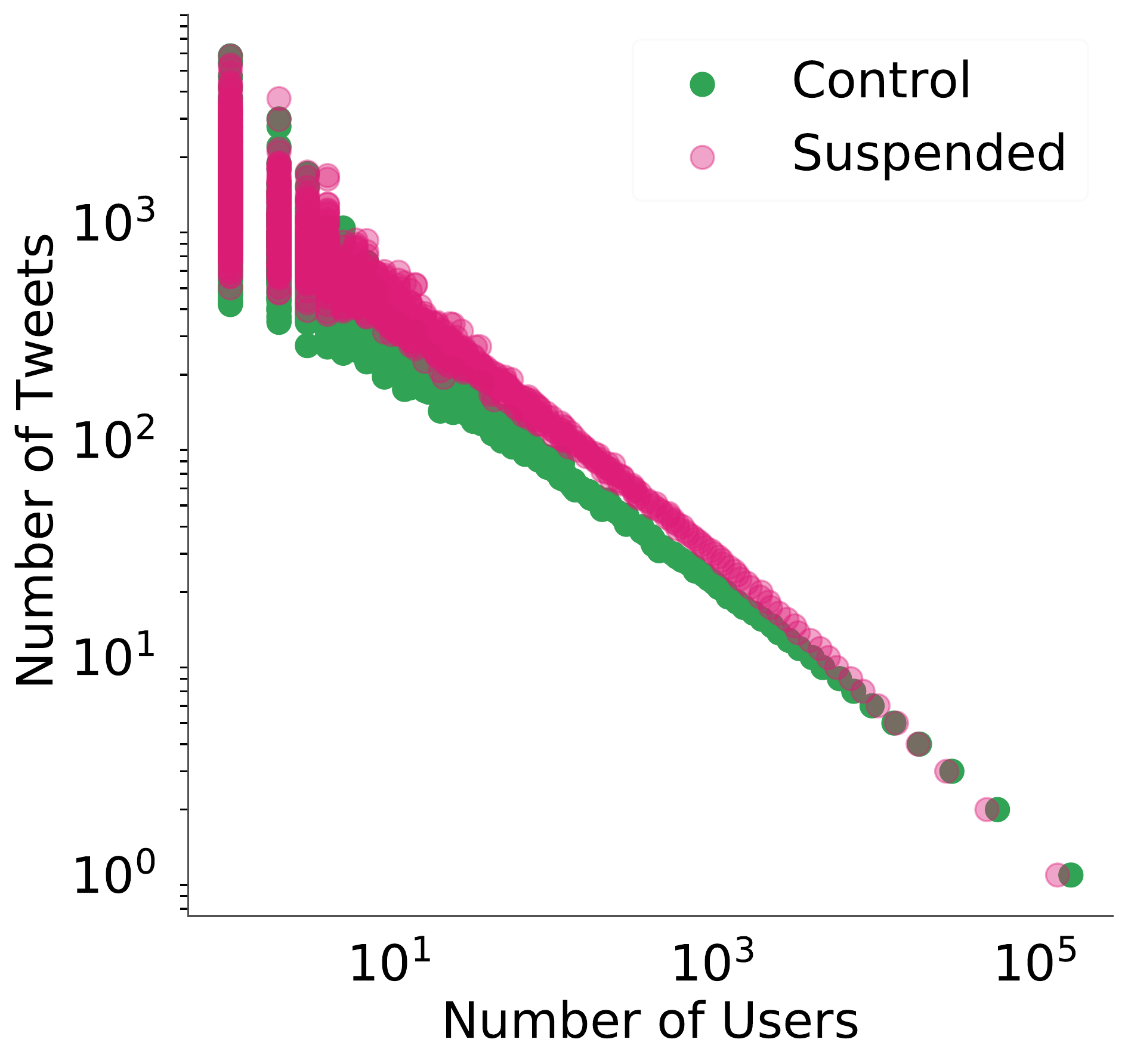}
      \label{fig:tweet_collection}
      \vspace{-1.8em}
      \captionof{figure}{Distribution of \# of users by \# of tweets.}
    \end{minipage}\hfill
    \begin{minipage}[t]{0.52\columnwidth}
  \vspace{-10em}
  \renewcommand{\arraystretch}{1.1}
      \centering
   \resizebox{\columnwidth}!{ \begin{tabular}[t]{lr}
      \textbf{Statistic} & \textbf{Value}\\
      \toprule
      \# Tweets & 240M\\
    \rowcollight \# Unique users & 21M\\
    \# Retweets & 173M\\
    \rowcollight \# Quote tweets & 60M\\
    \# $\mu$ tweets per user & 11\\
    \rowcollight \# Suspended users & 355K\\
    \# Tweets by sus. users & 7.2M\\
      \bottomrule
      \end{tabular}}
      \captionof{table}{Descriptive statistics of Twitter Dataset.}
      \label{tweet_collection}
  \end{minipage}\hfill
  \end{table}

\noindent
\textbf{Ethical Concerns.} Throughout our data collection, experiment design, and analysis process, we maintain ethical research standards~\cite{rivers2014ethical}. Hence, the accommodating academic institution's \textit{Institutional Review Board} exempted this project after a formal review. Following Twitter's terms of services guideline: (1) we use the Twitter API key only for passive data collection purposes, (2) we do not publish user-specific information, (3) we do not redistribute the data, and (4) we only share aggregated statistics and derived information to facilitate future work.

\section{Methods}
\subsection{RQ1: Inferring Suspension Factors}
\textbf{Twitter rules and policies.} 
To infer the plausible factors that explain suspension, we draw upon Twitter's rules, and policies for free and safe public discussion~\cite{twitter_rules_and_policies}. 
Twitter outlines three specific categories --- (1) safety, (2) privacy, (3) authenticity, each of which entails finer sub-categories on specific violating activities. 
We specifically focus on five sub-categories that are more likely to be enacted upon on election discussion: three from safety --- (1) hateful conduct, (2) abuse/harassment, (3) terrorism/violent extremism; two from authenticity --- (4) spamming, (5) civic integrity. The rest are either not largely relevant (i.e., copyright, nudity) or inferred from our data (i.e., impersonation - as we do not have user and tweet information for all the tweet, sensitive media content - as we do not crawl media content).\looseness=-1



\vspace{1em}
\noindent\textbf{Hateful Conduct and Offensive Behavior.} Several recent works aim to identify hateful and abusive activities in online platforms, which produced publicly available datasets and trained models~\cite{founta2018large, davidson2017automated}. However, as the distinction between abusive and violent language is ill-defined, they unified these categories. Similarly, we combine both abusive and violent tweets into one category - \textit{offensive}. We utilize an automatic hate speech and offensive language detection technique to detect hateful and offensive tweet, known as \textit{HateSonar}~\cite{davidson2017automated}. HateSonar is a logistic regression classifier that uses several text features (i.e., TF-IDF of word-grams, sentiment, etc.), which has been trained on manually labeled tweet corpus. We use a pre-trained HateSonar model to classify each tweet into three categories: (1) hateful, (2) offensive, and (3) normal.

\vspace{1em}
\noindent\textbf{Civic integrity.} Twitter has established strict rules to prevent users from ``manipulating or interfering in elections or other civic processes'', including ``posting and sharing misleading content''~\cite{twitter_civic_integrity}. To infer such violation, we utilize the posted hashtags and shared news website URLs. We curate a list of hashtags related to misinformation, propaganda, and conspiracy theories, borrowing from the work by~\cite{ferrara2020characterizing}, where they curated a list of conspiracy related hashtags. Additionally, we compile a list of biased and propaganda-spreader news websites based on a publicly available dataset from~\cite{factcheck_fake_site, politifact_fake_site}. If a tweet contains a hashtag or news article from our curated list, we consider it a violation of civic integrity policy. 

\vspace{1em}
\noindent\textbf{Spam.} Several previous works have identified spamming on Twitter, most of which consider both tweet content and user attributes for user-level classification. Here, we primarily infer spamming violations at tweet-level based on tweet content, for which we utilize a collection of spam keywords~\cite{benevenuto2010detecting}. However, to quantify spammers at the user-level, we also examine several account attributes (i.e., account age, tweet rate, etc.), which are most prominent for spammer detection~\cite{thomas2011suspended, yang2020scalable}. 

We note that the above-defined classification techniques are no match to Twitter's actual countermeasure mechanism. Rather, we posit these methods as high-precision approaches --- which utilize language models and keyword matching to avoid false-positives. The detected violations can be regarded as the lower-bound for actual ensued violations, only to increase with more comprehensive approaches.
\\
\vspace{-1.5em}
\subsection{RQ2: Political Ideology of Suspended Users}
We infer political leaning based on the political bias of the shared media outlets in the tweets. Similar to previous work on studying political ideology on Twitter~\cite{badawy2019characterizing}, we curate a list of ``politically inclined'' media outlets based on publicly available data from~\cite{allside_news_bias, media_news_bias}. Additionally, if a user retweets one of the presidential candidates without adding a quote, we consider it as ideological endorsement~\cite{ferrara2020characterizing}. 

\subsection{RQ3: Conversational topics and shared content} Twitter employs extensive countermeasure tools that consider a multitude of factors, features, and algorithms~\cite{twitter_measures}; which is beyond the scope of any third-party observer to reproduce. However, through the lens of Twitter's suspension policy, we can identify platform violators' targeted topics as a passive sensing mechanism for detecting online malice. Towards that, we contrast the conversational topics of \textsl{Suspended} and \textsl{Control} users. In particular, we consider (1) top uni-grams and bi-grams -- to infer the commonality of discussion language; (2) hashtags -- which are used for signaling and discoverability purposes~\cite{bruns2011use} and have been instrumental in several political and social movements~\cite{arif2018acting}. Additionally, Twitter is often used as an amplifier and fishing platform to disseminate news and multimedia content. Hence, we also examine the shared URL-domains to examine the online content platforms utilized by platform violators.

To examine the uniqueness across these dimensions, we use a generative text modeling method known as Sparse Additive Generative Models of Text, or SAGE~\cite{eisenstein2011sparse}. We use SAGE as a text differentiation technique where each class label or latent topic is endowed with a model of the deviation in log-frequency from a constant background distribution. We utilize SAGE to identify the highly used distinctive word-grams, hashtags, and URL-domains across \textsl{Suspended} and \textsl{Control} user's tweet corpus. \looseness=-1

\section{Results}
\begin{table}[t]
\sffamily
\centering
\footnotesize
\renewcommand{\arraystretch}{.97}
    \setlength{\tabcolsep}{2pt}
          \resizebox{\columnwidth}!{ 
    \begin{tabular}{lrrrrr}
    \textbf{Measure} &  \textbf{\Cs{}} &  \textbf{\Ct{}} & \textbf{\textsl{d}} & \textbf{\textsl{t}} & \textbf{\textsl{KS}}\\
    \midrule
    \rowcollight \multicolumn{6}{l}{Suspension rule (Safety)}\\
    \hspace{2mm} Hateful (\%) & 0.78 & 0.43 & 0.05 & 77.73*** & 0.003*** \\
    \hspace{2mm} Offensive (\%) & 6.45 & 5.21 & 0.05 & 91.62*** & 0.01*** \\
    \hdashline
    \rowcollight \multicolumn{6}{l}{Suspension rule (Authenticity)}\\
    \hspace{2mm} Civic Integrity (\%) & 0.40 & 0.30 & 0.02 & 31.46*** & 0.001*** \\
    \hspace{2mm} Spam (\%) & 0.56 & 0.38 & 0.03 & 46.83*** & 0.002*** \\
    \hdashline
    \rowcollight \multicolumn{6}{l}{Account Properties}\\
    \hspace{2mm} Active days & 964.0 & 1972.9 & -0.74 & -151.77*** & 0.24***\\
    \hspace{2mm} Tweets per Day &  33.5 & 20.82 & 0.25 & 50.20*** & 0.14***\\
    \hspace{2mm} Followers Count & 1491.5 & 3337.2 & -0.04  & -8.54*** & 0.11***\\
    \hspace{2mm} Friends Count & 1112.5 & 1263.7 & -0.03 & -6.74*** & 0.11***\\
    \hdashline
    \rowcollight \multicolumn{6}{l}{Political Ideology (\% Tweets)}\\
    \hspace{2mm} Left Leaning & 3.97 & 5.65 & -0.08 & -137.9*** & 0.02***\\
    \hspace{2mm} Right Leaning & 7.25 & 6.00 & 0.05 & 87.21*** & 0.01***\\
    \midrule
\end{tabular}}
    \caption{Summary of differences in quantitative measures across \Cs{} and \Ct{} users. We report average occurrences across matched clusters, effect size (Cohen's $d$), independent sample $t$-statistic, and $KS$-statistic. $p$-values are reported after Bonferroni correction (* $p$\textless0.05, ** $p$\textless0.01, *** $p$\textless0.001). }
\label{table:hypotests}
\end{table}

\subsection{RQ1: Inferring Suspension Reason}
In this subsection, we quantify the differences in the suspension rules between \textsl{Suspended} and \textsl{Control} users. We calculate effect size (Cohen's $d$) and use independent sample $t$-tests to evaluate statistical significance in the differences. We perform Koglomorov-Smirnov ($KS$) test to test against the null hypothesis that the distribution of suspension rules for the \textsl{Suspended} and \textsl{Control} users are drawn from the same distribution~\cite{sahaadvertiming}. We summarize these differences in Table \ref{table:hypotests}. 

\vspace{0.5em}
\noindent\textbf{Hateful Conduct and Offensive Behavior.} \textsl{Suspended} users are twice more likely to post hateful tweets than \textsl{Control} users ($t$=77.73, $p$\textless0.05). Also, \textsl{Suspended} users post more offensive tweet than \textsl{Control} users ($t$=91.62, $p$\textless0.05). These findings are in coherence with Twitter's suspension policy.

\vspace{0.5em}
\noindent\textbf{Civic Integrity and Spam.} We find that \textsl{Suspended} users are more likely to use hashtags related to conspiracy theories and share news from media sites with questionable authenticity ($t$=31.46, $p$\textless0.05). Also, \textsl{Suspended} users are 50\% more likely to post spam-tweets ($t$=46.83, $p$\textless0.05).

\vspace{0.5em}
\noindent\textbf{Account Properties.} We find significant difference in the active days between \textsl{Suspended} and \textsl{Control} users ($t$=-151.77, $p$\textless0.05). The \textsl{Control} users are, on average, roughly three years older than \textsl{Suspended} users. Contrastingly, these short-lived \textit{Suspended} users post 50\% more tweets compared to \textsl{Control} users ($t$=50.20, $p$\textless0.05). The \textsl{Suspended} users have less follower count, on average 100\% less than \textsl{Control} users ($t$=-8.54, $p$\textless0.05). However, both \textsl{Suspended} and \textsl{Control} users have similar friends count ($t$=-6.74, $p$\textless0.05). These findings resonate with previous works studying spamming and suspension on Twitter~\cite{thomas2011suspended, chowdhury2020twitter}, which identified that the rules violating users are generally short-lived, posts more tweet, and have a smaller follower base.  \looseness=-1

\subsection{RQ2: Political Ideology of Suspended Users}
In Table \ref{table:hypotests}, we observe $40$\% higher left-leaning tweets among \textsl{Control} users than \textsl{Suspended} users ($t$=-$137.9$, $p$\textless$0.05$). In contrast, we observe higher right-leaning tweets among \textsl{Suspended} users than \textsl{Control} users ($t$=$87.21$, $p$\textless$0.05$). Our finding shows that both left-leaning and right-leaning users engaged in violating Twitter's rules and policies, with $100$\% higher presence of right-leaning tweets among \textsl{Suspended} users than \textsl{Control} users.

\subsection{RQ3: Conversational topics and shared content}
\begin{table}
\sffamily
\fontsize{5}{5}\selectfont
\centering
\small
    \begin{tabular}{m{4.5em} m{6cm}} 
    \textbf{Category} & \textbf{Words}\\
    \midrule
    Control &  transistion, health care, amy coney barrett, flynn, senate, sidney powell, absentee, graham, judge, legal, rigged, mail, ballot, rudy giuliani, fraudulent\\
    \midrule
    Suspended & traitor, dumb, communist, biden family, idiot, seanhannity, fu*k, liar, stupid, breitbartnews, ukraine, treason, terrorist, evil, leftist\\
    \bottomrule
\end{tabular}
\caption{Highly used fifteen distinctive words per user-group obtained using SAGE.}
\label{table:distinct_words}
\vspace{-0.5em}
\end{table}

\vspace{0.5em}
\textbf{Word Usage.} In Table \ref{table:distinct_words}, we present the top 15 most distinctive words used in per group's tweet as obtained from the SAGE technique. We observe that \textsl{Control} users distinctly used words relating to different event-driven election-related topics (i.e., mail, ballot, rigged, fraudulent). However, we observe a large presence of swear words (i.e., idiot, dumb) and sensitive words (traitor, treason, terrorist) among \textsl{Suspended} users' unique words, which supports the higher hate-speech detection among suspended users.  

\begin{table}
\sffamily
\fontsize{5}{5}\selectfont
\centering
\small
    \begin{tabular}{m{4.5em} m{6cm}} 
    \textbf{Category} & \textbf{Hashtags}\\
    \midrule
    Control &  pentagon, bigtech, corruptkelly, quidproquo, doj, justicematters, climatechange, flipthesenate, michiganhearing, cnntapes\\
    \midrule
    Suspended & stevebannon, bidenfamilycorruption, warroompandemic, russia, hunterbidenemails, hunterbidenlaptop, democratsaredestroyingamerica, bidencrimesyndicate, chinajoe, chinabitchbiden\\
    \bottomrule
\end{tabular}
\vspace{-0.5em}
\caption{Highly used ten distinctive hashtags per user-group obtained using SAGE.}
\label{table:distinct_hashtags}
\end{table}

\vspace{0.5em}
\noindent\textbf{Hashtag Usage.} Table \ref{table:distinct_hashtags} presents the top 10 most distinctively used hashtags by \textsl{Suspended} and \textsl{Control} users. Similar to word usage, the unique hashtags among \textsl{Control} users were related to specific events (i.e., cnntapes, corruptkelly, etc) and general election issues (i.e., bigtech, climatechange). In contrast, distinct hashtags from \textsl{Suspended} users were mostly related to the defamation of Democratic presidential candidate Joe Biden (i.e., bidencrimesyndicate, chinajoe, chinabitchbiden) and issues related to his son Hunter Biden (i.e., hunterbidenemails, hunterbidenlaptops).  

\begin{table}
\sffamily
\fontsize{4.5}{4.5}\selectfont
\centering
\small
    \begin{tabular}{m{4.55em} m{6cm}} 
    \textbf{Category} & \textbf{Domain}\\
    \midrule
    Control &  democracydocket.com, buildbackbetter.gov, www.infobae.com, latimes.com, texastribune.com, citizensforethics.org, theatlantic.com, motherjones.com, nytimes.com, npr.org\\
    \midrule
    Suspended & usfuturenews.com, trumpsports.org, techfinguy.com, mostafadahroug.com, ovalofficeradio.com, wuestdevelopment.de, queenofsixteens.com, truenewshub.com, einnews.com, thefreeliberty.com\\
    \bottomrule
\end{tabular}
\vspace{-0.5em}
\caption{Highly shared ten distinctive URL-domain per user-group obtained using SAGE.}
\label{table:distinct_urls}
\end{table}

\vspace{0.5em}
\noindent\textbf{Shared Content.} In Table \ref{table:distinct_urls}, we show the top 10 distinct shared domain names per user group. Among the \textsl{Control} users, we observe the presence of few moderately neutral news outlets (i.e., nytimes.com, npr.org), independent political monitoring organization (i.e., democracydocekt.com, citizenforethics.org), and few left-leaning news outlets (theatlantic.com, motherjones.com). However, among \textsl{Suspended} users, we notice several heavily right-leaning non-mainstream news-propaganda sites (i.e., usfuturenews.com, ovalofficeradio.com, trunewshub.com, thefreeliberty.com). \looseness=-1

\section{Discussion and Conclusion}
\textbf{Implications.} Our study bears an implication in shedding light on the transparency about Twitter's content moderation policy. Although we cannot ascertain any quantitative estimation towards how far or through what means Twitter's rules followed, our study makes insightful findings of the statistically significant occurrences of hateful, offensive, and misinformative content among the users whose accounts were suspended after a while. These findings support theoretical, empirical, and anecdotal evidence about Twitter's moderation policies~\cite{twitter_rules_and_policies}, which had only gained significant attention since January 2021 when Twitter suspended the U.S. President Donald Trump's Twitter account owing to inciteful and unrest-provocative content~\cite{trump_ban}.

\vspace{0.5em}
\noindent\textbf{Limitations and Future Work.} Our Twitter data collection has potential biases as we initialize our seed keywords manually. While investigating plausible suspension reasons, we use simple, interpretable, and high-precision approaches - which are no match to Twitter’s complex and multi-faceted safeguard mechanisms. We do not infer the exact reason for suspension for individual users; rather, we quantify violations at tweet level. Future research can use causal inference methods like matching~\cite{saha2020causal} to minimize confounds and draw causal claims about why certain accounts were suspended. Moreover, we utilize several publicly available datasets that might suffer from biases.

We argue to situate our work as an initial step towards understanding malice, misinformation, and subsequent moderation related to the 2020 U.S. presidential election on online platforms. Our presented insights and the derived information can instigate further in-depth examination. For example, the shared news articles' content can be analyzed to understand the nature of propaganda news. To facilitate such research, we make these news URLs publicly available and other summary statistics ~\cite{support_materials_website}. Similarly, future works can investigate the dynamics of propaganda hashtags and news articles unique to suspended users to understand their impact and influence. Additionally, interactions among suspended users can be explored to identify potential coordination.

\noindent\textbf{Conclusion.} In this work, we perform a computational study to analyze Twitter's suspension policy situated in the context of the 2020 U.S. presidential election. We facilitate our work by collecting large-scale tweet dataset during the election period and subsequently identifying the suspended users. By designing a \textit{Case-Control} experimental study and devising high-precision classification approaches, we quantify associated factors related to the suspension. Additionally, we explore the political ideology and targeted topics of suspended users. We aim to motivate more rigorous and in-depth future works through our presented insights and shared datasets.

{
\fontsize{8.4pt}{8.4pt}
\selectfont\bibliography{reference,Mendeley1,Mendeley}}

\begin{thebibliography}{44}
\providecommand{\natexlab}[1]{#1}
\providecommand{\url}[1]{\texttt{#1}}
\providecommand{\urlprefix}{URL }
\expandafter\ifx\csname urlstyle\endcsname\relax
  \providecommand{\doi}[1]{doi:\discretionary{}{}{}#1}\else
  \providecommand{\doi}{doi:\discretionary{}{}{}\begingroup
  \urlstyle{rm}\Url}\fi

\bibitem[{Abu-El-Rub and Mueen(2019)}]{abu2019botcamp}
Abu-El-Rub, N.; and Mueen, A. 2019.
\newblock Botcamp: Bot-driven interactions in social campaigns.
\newblock In \emph{The World Wide Web Conference}, 2529--2535.

\bibitem[{AllSides(2021)}]{allside_news_bias}
AllSides. 2021.
\newblock {\url{https://www.allsides.com/media-bias/media-bias-ratings}}.

\bibitem[{Amleshwaram et~al.(2021)Amleshwaram, Reddy, Yadav, Gu, and
  Yang}]{amleshwaram2013cats}
Amleshwaram, A.~A.; Reddy, A.~N.; Yadav, S.; Gu, G.; and Yang, C. 2021.
\newblock CATS: Characterizing automation of Twitter spammers.

\bibitem[{Arif, Stewart, and Starbird(2018)}]{arif2018acting}
Arif, A.; Stewart, L.~G.; and Starbird, K. 2018.
\newblock Acting the part: Examining information operations within\#
  BlackLivesMatter discourse.
\newblock \emph{Proceedings of the ACM on Human-Computer Interaction} 2(CSCW):
  1--27.

\bibitem[{Badawy et~al.(2019)Badawy, Addawood, Lerman, and
  Ferrara}]{badawy2019characterizing}
Badawy, A.; Addawood, A.; Lerman, K.; and Ferrara, E. 2019.
\newblock Characterizing the 2016 Russian IRA influence campaign.
\newblock \emph{Social Network Analysis and Mining} 9(1): 31.

\bibitem[{Badawy, Ferrara, and Lerman(2018)}]{badawy2018analyzing}
Badawy, A.; Ferrara, E.; and Lerman, K. 2018.
\newblock Analyzing the digital traces of political manipulation: The 2016
  russian interference twitter campaign.
\newblock In \emph{2018 IEEE/ACM International Conference on Advances in Social
  Networks Analysis and Mining (ASONAM)}, 258--265. IEEE.

\bibitem[{Benevenuto et~al.(2010)Benevenuto, Magno, Rodrigues, and
  Almeida}]{benevenuto2010detecting}
Benevenuto, F.; Magno, G.; Rodrigues, T.; and Almeida, V. 2010.
\newblock Detecting spammers on twitter.

\bibitem[{Bessi and Ferrara(2016)}]{bessi2016social}
Bessi, A.; and Ferrara, E. 2016.
\newblock Social bots distort the 2016 US Presidential election online
  discussion.
\newblock \emph{First Monday} 21(11-7).

\bibitem[{Bias(2016)}]{twitter_bias_1}
Bias. 2016.
\newblock
  {\url{https://www.usatoday.com/story/tech/news/2016/11/18/conservatives-accuse-twitter-of-liberal-bias/94037802/}}.

\bibitem[{Bias(2020)}]{twitter_bias_2}
Bias. 2020.
\newblock
  {https://www.wired.co.uk/article/twitter-political-account-ban-us-mid-term-elections}.

\bibitem[{Bruns and Burgess(2011)}]{bruns2011use}
Bruns, A.; and Burgess, J.~E. 2011.
\newblock The use of Twitter hashtags in the formation of ad hoc publics.
\newblock In \emph{Proceedings of the 6th European consortium for political
  research (ECPR) general conference 2011}.

\bibitem[{Chowdhury et~al.(2020)Chowdhury, Allen, Yousuf, and
  Mueen}]{chowdhury2020twitter}
Chowdhury, F.~A.; Allen, L.; Yousuf, M.; and Mueen, A. 2020.
\newblock On Twitter Purge: A Retrospective Analysis of Suspended Users.
\newblock In \emph{Companion Proceedings of the Web Conference 2020}, 371--378.

\bibitem[{Congress-Hearing(2017)}]{congress_hearing}
Congress-Hearing. 2017.
\newblock
  {\url{https://www.govinfo.gov/content/pkg/CHRG-115shrg27398/pdf/CHRG-115shrg27398.pdf}}.

\bibitem[{Davidson et~al.(2017)Davidson, Warmsley, Macy, and
  Weber}]{davidson2017automated}
Davidson, T.; Warmsley, D.; Macy, M.; and Weber, I. 2017.
\newblock Automated hate speech detection and the problem of offensive
  language.
\newblock In \emph{Proceedings of the International AAAI Conference on Web and
  Social Media}, volume~11.

\bibitem[{Eisenstein, Ahmed, and Xing(2011)}]{eisenstein2011sparse}
Eisenstein, J.; Ahmed, A.; and Xing, E.~P. 2011.
\newblock Sparse additive generative models of text .

\bibitem[{Facebook-Update(2017)}]{facebook_update}
Facebook-Update. 2017.
\newblock
  {\url{https://about.fb.com/news/2017/09/information-operations-update/}}.

\bibitem[{FactCheck(2021)}]{factcheck_fake_site}
FactCheck. 2021.
\newblock
  {\url{https://www.factcheck.org/2017/07/websites-post-fake-satirical-stories/}}.

\bibitem[{Ferrara(2017)}]{ferrara2017disinformation}
Ferrara, E. 2017.
\newblock Disinformation and social bot operations in the run up to the 2017
  French presidential election.
\newblock \emph{arXiv preprint arXiv:1707.00086} .

\bibitem[{Ferrara et~al.(2020)Ferrara, Chang, Chen, Muric, and
  Patel}]{ferrara2020characterizing}
Ferrara, E.; Chang, H.; Chen, E.; Muric, G.; and Patel, J. 2020.
\newblock Characterizing social media manipulation in the 2020 US presidential
  election.
\newblock \emph{First Monday} .

\bibitem[{Founta et~al.(2018)Founta, Djouvas, Chatzakou, Leontiadis, Blackburn,
  Stringhini, Vakali, Sirivianos, and Kourtellis}]{founta2018large}
Founta, A.; Djouvas, C.; Chatzakou, D.; Leontiadis, I.; Blackburn, J.;
  Stringhini, G.; Vakali, A.; Sirivianos, M.; and Kourtellis, N. 2018.
\newblock Large scale crowdsourcing and characterization of twitter abusive
  behavior.
\newblock In \emph{Proceedings of the International AAAI Conference on Web and
  Social Media}, volume~12.

\bibitem[{Gil~de Z{\'u}{\~n}iga, Jung, and Valenzuela(2012)}]{gil2012social}
Gil~de Z{\'u}{\~n}iga, H.; Jung, N.; and Valenzuela, S. 2012.
\newblock Social media use for news and individuals' social capital, civic
  engagement and political participation.
\newblock \emph{Journal of computer-mediated communication} 17(3): 319--336.

\bibitem[{Im et~al.(2020)Im, Chandrasekharan, Sargent, Lighthammer, Denby,
  Bhargava, Hemphill, Jurgens, and Gilbert}]{im2020still}
Im, J.; Chandrasekharan, E.; Sargent, J.; Lighthammer, P.; Denby, T.; Bhargava,
  A.; Hemphill, L.; Jurgens, D.; and Gilbert, E. 2020.
\newblock Still out there: Modeling and identifying russian troll accounts on
  twitter.
\newblock In \emph{12th ACM Conference on Web Science}, 1--10.

\bibitem[{Le et~al.(2019)Le, Boynton, Shafiq, and
  Srinivasan}]{le2019postmortem}
Le, H.; Boynton, G.; Shafiq, Z.; and Srinivasan, P. 2019.
\newblock A postmortem of suspended Twitter accounts in the 2016 US
  presidential election.
\newblock In \emph{2019 IEEE/ACM International Conference on Advances in Social
  Networks Analysis and Mining (ASONAM)}, 258--265. IEEE.

\bibitem[{MediaBias(2021)}]{media_news_bias}
MediaBias. 2021.
\newblock {\url{https://mediabiasfactcheck.com/}}.

\bibitem[{Mueller-Report(2019)}]{mueller_report}
Mueller-Report. 2019.
\newblock {\url{https://www.justice.gov/storage/report.pdf}}.

\bibitem[{Olteanu et~al.(2015)Olteanu, Castillo, Diakopoulos, and
  Aberer}]{olteanu2015comparing}
Olteanu, A.; Castillo, C.; Diakopoulos, N.; and Aberer, K. 2015.
\newblock Comparing events coverage in online news and social media: The case
  of climate change.
\newblock In \emph{Proceedings of the International AAAI Conference on Web and
  Social Media}, volume~9.

\bibitem[{Olteanu et~al.(2014)Olteanu, Castillo, Diaz, and
  Vieweg}]{olteanu2014crisislex}
Olteanu, A.; Castillo, C.; Diaz, F.; and Vieweg, S. 2014.
\newblock Crisislex: A lexicon for collecting and filtering microblogged
  communications in crises.
\newblock In \emph{Proceedings of the International AAAI Conference on Web and
  Social Media}, volume~8.

\bibitem[{Politifact(2021)}]{politifact_fake_site}
Politifact. 2021.
\newblock
  {\url{https://www.politifact.com/article/2017/apr/20/politifacts-guide-fake-news-websites-and-what-they/}}.

\bibitem[{Rivers and Lewis(2014)}]{rivers2014ethical}
Rivers, C.~M.; and Lewis, B.~L. 2014.
\newblock Ethical research standards in a world of big data.
\newblock \emph{F1000Research} 3.

\bibitem[{Saha et~al.(2021)Saha, Liu, Vincent, Chowdhury, Neves, Shah, and
  Bos}]{sahaadvertiming}
Saha, K.; Liu, Y.; Vincent, N.; Chowdhury, F.~A.; Neves, L.; Shah, N.; and Bos,
  M.~W. 2021.
\newblock AdverTiming Matters: Examining User Ad Consumption for Effective Ad
  Allocations on Social Media.
\newblock In \emph{Proc. CHI}.

\bibitem[{Saha and Sharma(2020)}]{saha2020causal}
Saha, K.; and Sharma, A. 2020.
\newblock Causal Factors of Effective Psychosocial Outcomes in Online Mental
  Health Communities.
\newblock In \emph{Proceedings of the International AAAI Conference on Web and
  Social Media}, volume~14, 590--601.

\bibitem[{Schulz and Grimes(2002)}]{schulz2002case}
Schulz, K.~F.; and Grimes, D.~A. 2002.
\newblock Case-control studies: research in reverse.
\newblock \emph{The Lancet} 359(9304): 431--434.

\bibitem[{Thomas et~al.(2011)Thomas, Grier, Song, and
  Paxson}]{thomas2011suspended}
Thomas, K.; Grier, C.; Song, D.; and Paxson, V. 2011.
\newblock Suspended accounts in retrospect: an analysis of twitter spam.
\newblock In \emph{Proceedings of the 2011 ACM SIGCOMM conference on Internet
  measurement conference}, 243--258.

\bibitem[{Trump-Ban(2020)}]{trump_ban}
Trump-Ban. 2020.
\newblock
  {\url{https://blog.twitter.com/en_us/topics/company/2020/suspension.html}}.

\bibitem[{Twitter-Integrity(2021)}]{twitter_civic_integrity}
Twitter-Integrity. 2021.
\newblock
  {\url{https://help.twitter.com/en/rules-and-policies/election-integrity-policy}}.

\bibitem[{Twitter-Measures(2018)}]{twitter_measures}
Twitter-Measures. 2018.
\newblock
  {\url{https://blog.twitter.com/en_us/topics/company/2018/how-twitter-is-fighting-spam-and-malicious-automation.html}}.

\bibitem[{Twitter-Policy(2021)}]{twitter_election_policy}
Twitter-Policy. 2021.
\newblock
  {\url{https://blog.twitter.com/en_us/topics/company/2020/2020-election-changes.html}}.

\bibitem[{Twitter-Safety(2021)}]{twitter_safety}
Twitter-Safety. 2021.
\newblock
  {\url{https://blog.twitter.com/en_us/topics/company/2018/how-twitter-is-fighting-spam-and-malicious-automation.html}}.

\bibitem[{Twitter-Update(2018)}]{twitter_update}
Twitter-Update. 2018.
\newblock
  {\url{https://blog.twitter.com/en_us/topics/company/2018/2016-election-update.html}}.

\bibitem[{TwitterRules(2021)}]{twitter_rules_and_policies}
TwitterRules. 2021.
\newblock {\url{https://help.twitter.com/en/rules-and-policies/twitter-rules}}.

\bibitem[{Website(2021)}]{support_materials_website}
Website. 2021.
\newblock {\url{https://sites.google.com/view/us-election20-twitter-suspend}}.

\bibitem[{Woolley and Howard(2018)}]{woolley2018computational}
Woolley, S.~C.; and Howard, P.~N. 2018.
\newblock \emph{Computational propaganda: political parties, politicians, and
  political manipulation on social media}.
\newblock Oxford University Press.

\bibitem[{Yang et~al.(2020)Yang, Varol, Hui, and Menczer}]{yang2020scalable}
Yang, K.-C.; Varol, O.; Hui, P.-M.; and Menczer, F. 2020.
\newblock Scalable and generalizable social bot detection through data
  selection.
\newblock In \emph{Proceedings of the AAAI Conference on Artificial
  Intelligence}, volume~34, 1096--1103.

\bibitem[{Zannettou et~al.(2019)Zannettou, Caulfield, De~Cristofaro,
  Sirivianos, Stringhini, and Blackburn}]{zannettou2019disinformation}
Zannettou, S.; Caulfield, T.; De~Cristofaro, E.; Sirivianos, M.; Stringhini,
  G.; and Blackburn, J. 2019.
\newblock Disinformation warfare: Understanding state-sponsored trolls on
  Twitter and their influence on the web.
\newblock In \emph{Companion proceedings of the 2019 world wide web
  conference}, 218--226.

\end{thebibliography}
\balance{}

\end{document}